\documentclass[10pt]{revtex4}
\usepackage[dvips]{graphicx}
\usepackage{amsthm,latexsym,amssymb,amsfonts,array}
\begin{document}

\title{The regular cosmic string in Born-Infeld gravity}

\author{Rafael Ferraro $^{1,\, 2}$  and Franco Fiorini $^1$}

\address{$^1$ Instituto de Astronom\'\i a y F\'\i sica del
Espacio,Casilla de Correo 67, Sucursal 28, 1428 Buenos Aires,
Argentina.\\ $^2$ Departamento de F\'\i sica, Facultad de Ciencias
Exactas y Naturales, Universidad de Buenos Aires, 1428 Buenos Aires,
Argentina.}

\email{ferraro@iafe.uba.ar,franco@iafe.uba.ar}


\vskip2cm

\begin{abstract}
It is shown that Born-Infeld gravity --a high energy deformation of
Einstein gravity-- removes the singularities of a cosmic string. The
respective vacuum solution results to be free of conical singularity
and closed timelike curves. The space ends at a minimal circle where
the curvature invariants vanish; but this circle cannot be reached
in a finite proper time.
\end{abstract}

\maketitle

\vskip1cm

We shall investigate high energy deformations of General Relativity
(GR) based on the following three guiding principles:

\begin{description}

    \item[a)] The theory must reduce to General Relativity in the low
    field limit.

    \item[b)] The spacetime dynamics must be described by second order equations.

    \item[c)] It must provide a proper treatment in order to avoid singularities.
\end{description}

Many deformations can accomplish the condition (a), but just a few
of them satisfy (b).\footnote{For instance, the so called ``$f(R)$"
theories \cite{fr}, lead to fourth order equations. They can be
reformulated as scalar-tensor theories to give second order
equations; however this procedure results in violations of the
equivalence principle \cite{Eq,olmo}. Moreover, vacuum singular GR
solutions remain as solutions of smooth $f(R)$ theories, so
threatening the requirement (c).} Lovelock's extension of GR does it
\cite{love}, but it only departs from GR for dimensions bigger than
four. In turn, a few candidates were proposed following the
guidelines (a) and (c) (see e.g. the non-exhaustive list
\cite{deser3}-\cite{Nieto}). Remarkably, there is no conceptual
framework accommodating the three requirements. To face the item (c)
we will consider one of the prototypes of a GR singular structure:
the circular symmetric vacuum solution in (2+1) dimensions
\cite{star,3D}, or its (3+1)-dimensional analogue, the cosmic string
\cite{4D}
\begin{equation}
ds^2=d(t+4 J\,\theta)^2-d\rho^2-(1-4\mu)^2\, \rho^2\,
d\theta^2-dz^2. \label{cosmon}
\end{equation}
In (2+1) dimensions ($z$ is absent), this metric solves the Einstein
equations for $T^{00}=\mu\, \delta(x,y)$ and $T^{0i}=(J/2)\,
\epsilon^{ij}\, \partial_j \delta(x,y)$ (Cartesian coordinates $x,
y$ are defined as usual, and we use hereafter $G=1$). So the
solution looks as a particle of mass $\mu$ and spin $J$ at the
origin \cite{3D}. However, no gravitational field surrounds the
particle since the metric is manifestly flat, as is typical of
vacuum solutions to Einstein equations without cosmological constant
in (2+1)-dimensions. Actually the particle shows itself by means of
topological properties: i) the space displays a conical singularity,
because the mass $\mu$ amounts a deficit angle $8\pi\mu$; ii) the
Minkowskian time $t'\equiv t + 4 J\,\theta$ suffers a jump $8\pi\,
J$ when the circle is completed. Added to this singular structure,
the {\it cosmon} --as named in Ref.~\cite{deser4}-- also exhibits
violation of causality because closed timelike curves (CTC) occur.
In fact, the interval (\ref{cosmon}) on the closed curves of
constant $(t, \rho, z)$ is
\begin{equation}
ds^2=\left(\frac{16 J^2}{M^2}-\rho^2\right)\, M^2\, d\theta^2\, ,\
\ \ \ \ M \equiv 1-4\mu, \label{CTC}
\end{equation}
which means that curves with radio $\rho<4J/M$ are CTC. This
unpleasant feature was avoided in Ref.~\cite{deser4} by resorting to
proper boundary conditions. However, we will show that the entire
singular structure of cosmons is prevented by a high energy
deformation of GR developed in Refs.~\cite{Nos}-\cite{nos4}.

As the starting point for describing the gravitational field, we
will propose a {\it determinantal} Born-Infeld (BI) Lagrangian
density \cite{nos4}
\begin{equation}
\mathcal{L}_{\mathbf{BIG}}=-\lambda/(16\,\pi)\ \Big(\sqrt{|g_{\mu
\nu }-2\lambda ^{-1}F_{\mu \nu}|}-\sqrt{|g_{\mu \nu }|}\Big),
\label{acciondetelectro}
\end{equation}
where $|\ |$ stands for the absolute value of the determinant. The
structure (\ref{acciondetelectro}) resembles the one used by Born
and Infeld in their non linear electrodynamics \cite{BI}. In our
case, the field tensor $F_{\mu \nu}$ will depend on the dynamical
variable describing the gravitational field. The constant
$\lambda$ in Eq.~(\ref{acciondetelectro}) has the units of $F_{\mu
\nu}$ and controls the low energy limit alluded in the requirement
(a). In fact, since
\begin{equation}
\sqrt{|I-2\lambda^{-1}F|}=1-\lambda^{-1}Tr(F)+\mathcal{O}(\lambda^{-2}),
\end{equation}
then the Lagrangian (\ref{acciondetelectro}) goes to
\begin{equation}
\mathcal{L}_{\mathbf{G}}\ =\ 1/(16\,\pi)\ \sqrt{|g_{\mu \nu }|}\
Tr(F), \label{accionbaja}
\end{equation}
when $F_{\mu \nu}$ takes values much smaller than $\lambda$. So,
Born-Infeld gravity will reduce to GR if $Tr(F)\,=\,-R$. On the
other hand, the requirement (b) implies that the quantities involved
in the Lagrangian (\ref{acciondetelectro}) should depend just on the
dynamical variables and their first derivatives. This condition
could seem to be out of tune with GR, since the curvature scalar
depends on the second derivatives of its dynamical field (the
metric). However, there is a beautiful way of circumventing this
blind alley, by moving out towards an equivalent formulation of GR
which depends just on first derivatives of its dynamical variables.
Indeed, Einstein theory can be rephrased in a spacetime possessing
absolute parallelism \cite{Hehl2} where the {\it vielbein}
$\{e^{a}(x)\}$ --a set of 1-forms-- plays the role of the
gravitational potentials, while the metric $\mathrm{g}(x)=\eta
_{ab}\ e^{a}(x)\otimes e^{b}(x)$ is a subsidiary field. The absolute
parallelism is ruled by the Weitzenb\"{o}ck connection $\Gamma _{\mu
\nu }^{\lambda }\equiv e_{a}^{\ \lambda }\,e_{\ \mu,\,\nu }^{a}$,
where $\{e_{a}^{\ \lambda}(x)\}$ is the vector basis dual to the
{\it vielbein}. This connection is metric compatible and
curvatureless: Weitzenb\"{o}ck spacetime is flat, but it possesses
torsion $T_{\ \ \mu \nu }^{\rho }\, =\, e_{\ a}^{\rho }\,(e_{\
\nu,\,\mu }^{a}-e_{\ \mu,\, \nu }^{a})$ --i.e., $T^{a}=de^{a}$--,
which is the agent where the gravitational degrees of freedom are
encoded. The equivalence between both pictures reveals itself when
one discovers that the Levi-Civita curvature scalar $R$ can be
rewritten as
\begin{equation}
4\, R[e^a]\ =\ 2\,W_1\,-\,W_2\,-\,4\,W_3\, +\,8\,
\mathrm{e}^{-1}\,(\mathrm{e}\,T^{\mu\, \, \rho}_{\ \,
\mu})_{,\,\rho}\ , \label{equiv}
\end{equation}
where $\mathrm{e}=\sqrt{|\mathrm{g}|}$ is the determinant of the
matrix $e^{a}_{\,\,\mu}$, and the invariants $W_{i}$ are
\begin{equation}
W_{1}=T^{\mu\nu}_{\ \ \ \rho}\,T^{\rho}_{\ \ \mu\nu}\, ,
\hspace{0.25cm}W_{2}=T^{\ \ \mu\nu}_{\rho}\,T^{\rho}_{\ \ \mu\nu}\,
, \hspace{0.25cm}W_{3}=T^{\rho\nu}_{\ \ \ \rho}\,T^{\mu}_{\ \
\mu\nu}\, .
\end{equation}
Remarkably, the advantage of the teleparallel picture of gravity is
that the total divergence term in Eq.~(\ref{equiv}) can be ruled out
to work just with a first order Lagrangian $\mathcal{L}(e^a,
\partial e^a)$, which will preserve the dynamical content of
Einstein theory anyway. This distinctive feature makes
Weitzenb\"{o}ck spacetime a privileged geometric structure to
formulate modified theories of gravitation, since it guarantees that
any modified Lagrangian in this language will assure second order
field equations. This mechanism is essential to satisfy the
requirement (b), otherwise fourth order differential equations would
be obtained. Summarizing, we are building a theory of gravity where
the fundamental piece is the set of 2-forms $T^{a}=de^{a}$ (notice
the similarity with Yang-Mills theories), so guaranteeing the
requirement (b). To accomplish the requirement (a) we should demand
that
\begin{equation}
4\,Tr(F)\ = \ -2\,W_1\,+\,W_2\,+\,4\,W_3\ \equiv\ 4\, S_{\rho }^{\ \
\mu \nu }\, T^{\rho}_{\ \ \mu\nu},\label{condition}
\end{equation}
where
\begin{equation}
S_{\rho }^{\ \ \mu \nu }=-\frac{1}{4}\,\left(T_{\ \ \ \rho }^{\mu
\nu }-T_{\ \ \ \rho }^{\nu \mu }-T_{\rho }^{\,\ \mu \nu
}\right)+\frac{1}{2}\,\left(\delta _{\rho }^{\mu }\ T_{\ \ \ \theta
}^{\theta \nu }-\delta _{\rho }^{\nu }\ T_{\ \ \ \theta }^{\theta
\mu }\right). \label{tensorS}
\end{equation}
The more general tensor $F_{\mu \nu }$ fulfilling the condition
(\ref{condition}) reads $F_{\mu \nu }=\alpha\, S_{\mu \lambda \rho
}\, T_{\nu }^{\,\,\,\lambda \rho }+\beta\, S_{\lambda \mu \rho }\,
T_{\,\,\,\,\,\nu }^{\lambda \,\,\,\,\rho }$, $\alpha+\beta=1$. For
simplicity we will set $\beta=0$; thus, in four dimensions the
theory reduces to
\begin{equation}
\mathcal{L}_{\mathbf{BIG}}[e^a]=-\lambda/(16\,\pi)\!
\left(\sqrt{|g_{\mu \nu }-2\lambda ^{-1}S_{\mu \lambda \rho }\,
T_{\nu }^{\,\,\,\lambda \rho }|}- \sqrt{|g_{\mu \nu }|}\right).
\label{acciondet}
\end{equation}

\begin{figure*}[h]
\centering \includegraphics[scale=.25]{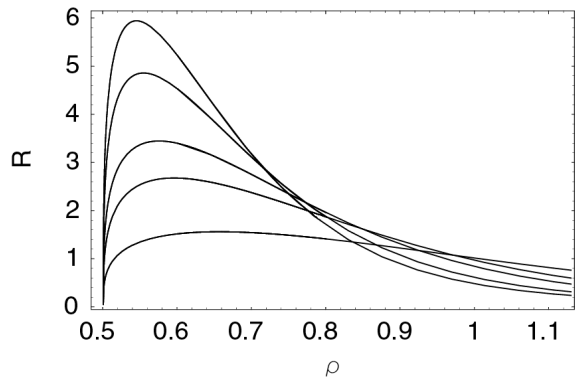}
\caption{\footnotesize{Scalar curvature $R(\rho)$ for $J/M=1/8$ in
units of length (so, $R$ vanishes at $\rho_{o}=1/2$). Following the
maximum of the curves from bottom to top, it is $\lambda=1, 3, 5,
10, 15$ in units of inverse square length. }}
 \label{fig1}
\end{figure*}

Let us now analyze the requirement (c). We will show that BI gravity
(\ref{acciondet}) prevents the singular structure of the GR cosmon
metric (\ref{cosmon}). First of all, we remark that the cosmon was
endowed with spin $J$ by considering the coordinate $t=t'-4J \theta$
as the continuous time in the place of the Minkowskian time $t'$; in
(2+1) dimensions, this procedure led to a non-zero $T^{0i}$ at the
origin. The mass $\mu$ came from a similar procedure. In BI gravity,
instead, the equations determine not the metric but the vielbein;
they contain more integration constants than GR does. In particular,
$J$ and $M\equiv 1-4\mu$ will appear as labels for the family of
cylindrically symmetric vielbeins solving the dynamical equations.
We found that the solution of the motion equations is given by the
vielbein \cite{nos4}
\begin{equation}
e^0\, =\, d(t+4J\theta)    \, ,\ \ \ \  e^1\, =\, Y(\rho)\,
d\rho\, ,\ \ \ \  e^2\, =\, \rho\, M\, d\theta\, ,\ \ \ \ e^3\,
=\, dz,\label{vielbein}
\end{equation}
where the function $Y(\rho)$ solves the cubic equation
\begin{equation}
Y^2(\rho)-Y^3(\rho)=-\frac{16\, J^2}{\lambda
M^2}\Big(\rho^2-\frac{16\, J^2}{M^2}\Big)^{-2}. \label{rela}
\end{equation}
The vielbein (\ref{vielbein}) implies the metric
\begin{equation}
ds^2=d(t+4 J\,\theta)^2-Y^2(\rho)d\rho^2-M^2\, \rho^2\,
d\theta^2-dz^2, \label{cosmonbi}
\end{equation}
then the function $Y(\rho)$ constitutes the sole difference between
GR and BI gravity. GR is recovered when $\lambda\rightarrow\infty$,
as expected from the low energy limit (\ref{accionbaja}). In fact,
Eq.~(\ref{rela}) says that $Y\rightarrow 1$ and the metric
(\ref{cosmonbi}) acquires the flat form of Eq.~(\ref{cosmon}). For
finite values of $\lambda$ and far away from the string, the low
field regime is also restituted. Nevertheless, the BI cosmic string
is a geometry with a $J$-depending non-constant scalar curvature
$R=-2Y'/(\rho Y^3)$ depicted in Figure \ref{fig1} for several values
of $\lambda$.

According to Eq.~(\ref{rela}), $Y$ goes to infinity as $\rho$
approaches the value $\rho_{o}=4J/M$. At this minimal circle, $R$
and other curvature invariants such as $R^{\mu\nu}R_{\mu\nu}$ and
$R^\lambda_{\,\,\ \mu\nu\rho}\, R_{\lambda}^{\,\,\ \mu\nu\rho}$
become zero. Thus, the spacetime is flat not only far from the
string (GR region) but at $\rho_{o}$ as well. It is easy to verify
that the proper time to reach $\rho_{o}$ diverges \cite{nos4}, which
means that BI gravity avoids the conical singularity. Moreover, the
CTC's are prevented too. Even though the interval on the closed
curves of constant $(t, \rho, z)$ does not differ from the one of
Eq.~(\ref{CTC}) --since $\rho$ is constant, then $Y(\rho)$ does not
play any role--, however CTC's are forbidden because the lower bound
of the radial coordinate $\rho$ implies that the interval
(\ref{CTC}) is negative definite. In this way, the same mechanism
responsible for the taming of the conical singularity in BI gravity
also provides a natural chronological protection.

\medskip

It is worth mentioning that the coordinate change $d\xi=Y(\rho)\,
d\rho$ in the vielbein (\ref{vielbein}) allows to regard the curved
geometry (\ref{cosmonbi}) as a space of a variable deficit angle
ranging from $8\pi\mu$ at spatial infinity, to $2\pi$ at $\rho_0$.
This feature might have important observational implications on the
lensing effect. As another remarkable physical consequence, BI
gravity seems to forbid the possibility of packing energy in
arbitrarily small regions. Differing from GR, any junction of this
vacuum solution with an inner solution has to be made at a radius
bigger than $\rho_o=4J/M$. This property is another manifestation of
the Born-Infeld regularization program. As a final remark, we
emphasize that the BI determinantal gravity (\ref{acciondet})
departs from a mere ``$f(T)$" theory even at the lowest order in
$\lambda^{-1}$ ($T$ stands for the invariant $S_{\rho }^{\ \ \mu \nu
}\, T^{\rho}_{\ \ \mu\nu}$). This feature is essential for smoothing
the cosmic string singularities, since the solution has $T=0$
\cite{Nos2}.

\medskip

\section*{Acknowledgments}This work was supported by CONICET and Universidad de Buenos
Aires.



\vskip1cm

\begin{thebibliography}{99}


\bibitem{fr}Capozziello S, De Laurentis M and Faraoni V 2009 A bird's eye view of
$f(R)$-gravity {\it Preprint} gr-qc/0909.4672
\bibitem{Eq}Brans C H 2005 The root of scalar-tensor theory: an approximate
history {\it Preprint} gr-qc/0506063
\bibitem{olmo}Olmo G J 2007 {\it Phys. Rev. Lett.} \textbf{98} 061101
\bibitem{love}Lovelock J 1971 {\it J. Math. Phys.} \textbf{12} 498
\bibitem{deser3}Deser S and Gibbons G W 1998 {\it Class. Quantum Grav.} \textbf{15} 35
\bibitem{Fein}Feingenbaum J A 1998 {\it Phys. Rev.} D \textbf{58} 124023
\bibitem{Comelli}Comelli D and Dolgov A 2004 {\it JHEP} \textbf{0411} 062
\bibitem{Tek2} G\"ull\"u I, \c{S}i\c{s}man T \c{C} and Tekin B 2010 {\it Phys. Rev.} D \textbf{81} 104018
\bibitem{Maximo}Ba\~nados M and Ferreira P 2010 {\it Phys. Rev. Lett.} \textbf{105} 011101
\bibitem{Nieto}Nieto J A 2004 {\it Phys. Rev.} D \textbf{70} 044042
\bibitem{star}Staruszkiewicz A 1963 {\it Acta. Phys. Polon.} \textbf{24} 734
\bibitem{3D}Deser S, Jackiw R and 't Hooft G 1984 {\it Ann. Phys.} \textbf{152}  220
\bibitem{4D}Gott III J R 1985 {\it Ap. J.} \textbf{288} 422
\bibitem{deser4}Deser S, Jackiw R and 't Hooft G 1992 {\it Phys. Rev. Lett.} \textbf{68} 267
\bibitem{Nos}Ferraro R and Fiorini F 2007 {\it Phys. Rev.} D \textbf{75} 084031
\bibitem{Nos2}Ferraro R and Fiorini F 2008 {\it Phys. Rev.} D \textbf{78} 124019
\bibitem{nos4}Ferraro R and Fiorini F 2010 {\it Phys. Lett.} B
\textbf{692} 206
\bibitem{BI}Born M and Infeld L 1934 {\it Proc. R. Soc.} A \textbf{144} 425
\bibitem{Hehl2}Hehl F W, McCrea J D, Mielke E W and Ne'eman Y 1995 {\it Phys. Rep.} \textbf{258} 1



\end{thebibliography}
\end{document}